\documentclass{article}
\usepackage{natbib}
\usepackage[french,spanish,english]{babel}
\usepackage{amsmath}
\usepackage{amssymb}
\usepackage{amsthm}
\usepackage[pdftex]{graphicx}
\usepackage{tabularx}
\usepackage{hyperref}
\usepackage[utf8]{inputenc}

\usepackage{a4wide}
\newcommand{\fah}{F\aa hr\ae us }
\newcommand{\fahNoS}{F\aa hr\ae us}
\newcommand{\gp}{\dot{\gamma}}

\newcommand{\gpmoy}{\left<\dot{\gamma}\right>}

\begin{document}

\title{Murray's law revisited:\\ Qu\'emada's fluid model and fractal trees.}


\author{Baptiste Moreau$^1$ and Benjamin Mauroy$^{2,*}$\\\\
\small \phantom{}$^1$ Laboratoire MSC, UMR CNRS 7057, Universit\'e Paris 7 Denis Diderot, France.\\
\small \phantom{}$^2$ Laboratoire JA Dieudonn\'e, UMR CNRS 7351, Universit\'e de Nice-Sophia Antipolis, France.\\
\small \phantom{}$^*$ corresponding author (\href{benjamin.mauroy@unice.fr}{benjamin.mauroy@unice.fr})}

\date{}

\maketitle

\begin{abstract}
We investigate how Murray's law is affected by fluids whose viscosity depends monotonously on shear rates, such as blood. Our study shows that Murray's original law also applies to such fluids, as long as they did not undergo phase separation effect. When \fahNoS-like phase separation effects occur, we derive an extended version of Murray's law. Finally, we study how these new laws affect the optimal geometries of fractal trees to mimic an idealized arterial network. Our analyses are based on Qu\'emada's fluid model, but our approach is very general and apply to most fluids with shear dependent rheology.
\end{abstract}


\section{Introduction}

\cite{murray_physiological_1926} proposed the first law for the optimal design of blood vessels, based on a trade-off between the power needed to make blood circulate in the vessel and the metabolic power needed to maintain blood. His work was based on the original researches developed by \cite{hess_prinzip_1914}. The law is formulated using Poiseuille's regime in cylindrical vessels,
thus it accounts for viscous effects only
and neglects any perturbations due to fluid divergence or convergence at the bifurcations points.
Blood flow rate is assumed constant to mimic for the fact that the vessels have to feed a downstream organ whose needs are independent on the vessels geometry. 
The optimal configuration corresponds to the blood flow rate being proportional to the cube power of the radius of the blood vessel. The well known corollary for a bifurcation states that the cube radius of the parent vessel equals the sum of the cube radii of the daughter vessels.
Murray's law is independent on the amount of blood flow rate, thus it does not depend on the functioning regime of the downstream organ, at least in the limit of its hypotheses.
In the seventies,  \cite{zamir_optimality_1976} extended this law to account for bifurcation angles and he expressed Murray's law in term of wall shear stress being constant independently on the vessel size \citep{zamir_shear_1977}. 
While blood arterial 
mid-level 
circulation meets conditions for Murray's law, this is not the case for 
the larger vessels, where inertia and/or turbulence occur \citep{uylings_optimization_1977}, and for
microcirculation, where wall shear stress is decreasing with the sizes of the vessels 
\citep{sherman_connecting_1981}. 
Different hypotheses were developed to explain the shifts to Murray's law observed with microcirculation, for example \cite{taber_optimization_1998} proposed to add an energy cost related to smooth muscles. 
Later, 
\cite{alarcon_design_2005} used semi-empirical laws from 
\cite{pries_resistance_1994} 
and showed that wall shear stress behavior in microcirculation can be explained by the \fahNoS-Lindvquist effect. The \fahNoS-Lindvquist effect is a phase separation effect occurring in small blood vessels that makes blood viscosity become a non monotonous function of vessel radius \citep{fahraeus_suspension_1929, pries_design_1995, fung_biomechanics_1997, pries_blood_2008}.
Similarly, studies were made to understand why Murray's law does not apply in large blood vessels, where inertia and turbulence play an important role. These studies were based on modified formulation of the power associated to fluid circulation, shifting the radius exponent from $3$ in laminar case down to $2.33$ for fully turbulent flow \citep{uylings_optimization_1977}. Finally, generalization of Murray's law has also been developed to account for the non-local properties of tree structures and extends the predictions using empirical exponents based on the fractal nature of the biological networks \citep{zhou_design_1999, kassab_scaling_2007}. These exponents aggregate rich information from the physiological configuration, amongst which the non linear behavior of blood viscosity. These last studies should however be considered with care, since the empirical data injected in the model is actually dependent on the variables that are optimized, which might bring a bias in the predictions. 

Another step to fully extend Murray's law to blood circulation is now to add the dependence of vessel blood viscosity on flow amplitude. Indeed, blood is a 
shear-thinning
fluid and its local viscosity depends on the local shear rate in the vessel. Local shear rate is a function of both blood flow amplitude in the vessel and vessel radius. Thus, in addition to red blood cells volume fraction and vessel radius, equivalent viscosity in blood microcirculation is expected to be also dependent on the shear rates in the vessel. 

First we show that if \fah effect does not occur, the fluid equivalent viscosity in a vessel is dependent on mean shear rate and hematocrit in the branch only. If \fah effect occurs, then we show that equivalent viscosity becomes also dependent on vessel radius. Then we apply Murray's optimal design to both cases with and without \fah effect and derive corresponding laws for optimal configuration of a vessel and of a bifurcation. Finally, we study how these new laws affect optimal geometries of fractal tree structures, using inherent properties for mean shear rates variation inside a fractal tree.

\section{Mathematical model}
 
Since Murray's law was formulated in the frame of the cardiovascular system, we chose to work with 
a model that mimics blood rheology.
However, blood exhibits a complex thixotropic behavior with a yield stress to overcome for blood to flow \citep{merrill_rheology_1969,bureau_rheological_1980,apostolidis_modeling_2014}. Moreover, blood rheology is not only affected by red blood cells concentration but also by its inner composition in proteins, such as fibrinogen concentration \citep{merrill_yield_1969}. We did not want to account for such refined behaviors of blood rheology since our study focused on the interaction between Murray's optimization process and a shear rate-dependent rheology. Thus we chose to work with Qu\'emada's fluid model
which tightly fits our needs and were initially proposed for modeling blood rheology.
Those fluids are well documented \citep{quemada_towards_1984} and have been intensively studied and validated as good approximations for blood modeling \citep{cokelet_rheology_1987, neofytou_comparison_2004, marcinkowska-gapinska_comparison_2007,sriram_non-newtonian_2014}. The viscosity in Qu\'emada's fluid model depends on the local shear rate $\gp$ and on the local red blood cells volume fraction $H$. Complete details about Qu\'emada's model are given in section I of supplementary materials. 
Notice that other models, such as Casson's model could also have been used in this study \citep{casson_flow_1959}.

\begin{figure}[h!]
\centering
\includegraphics[height=4.5cm]{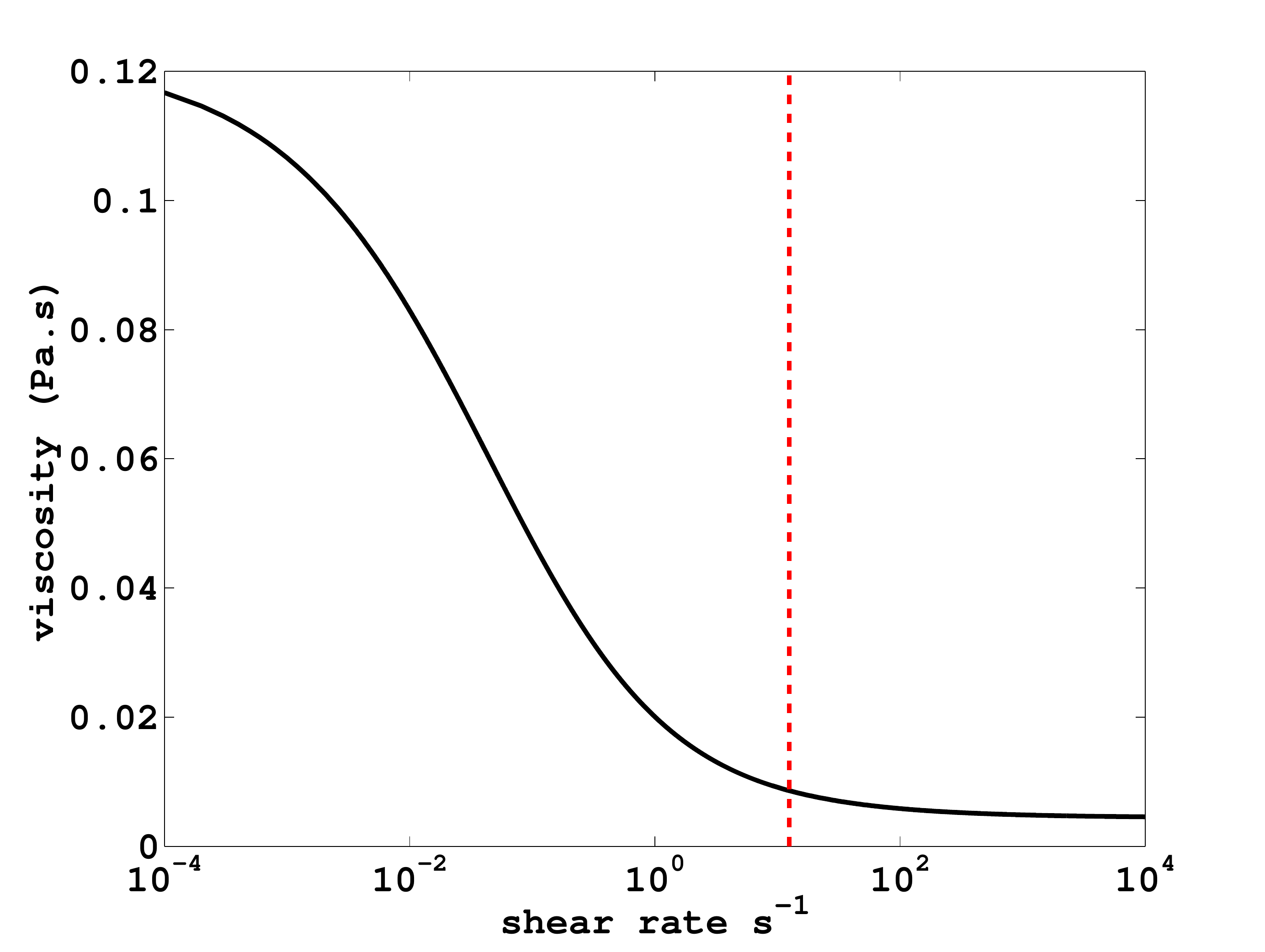}
\caption{Qu\'emada's model of viscosity \citep{quemada_towards_1984, cokelet_rheology_1987}. Black curve: viscosity dependence on the shear rate (blood case, $H = 0.45$). Red dashed line: optimal mean shear rate predicted by the model for blood large circulation $\gpmoy_{noF}^* \sim 12.5 \ s^{-1}$.}
\label{viscosityvs}
\end{figure}

In Qu\'emada's model, the dependence of fluid viscosity on shear rates divides into three parts, as shown on figure \ref{viscosityvs} (blood case): a plateau of high viscosity for low shear rates (blood: $<10^{-3} \ s^{-1}$); one sharp decrease of viscosity for "medium-ranged" shear rates (blood: from $10^{-3} \ s^{-1}$ to $1 \ s^{-1}$);  a plateau of low viscosity for high shear rates (blood: $> 1 \ s^{-1}$).

We assume the vessels to be cylindric and the fluid to be at low regime, axi-symmetric and fully developed in all the vessels. 
Blood flow rate through a vessel is assumed constant in order to mimic a downstream organ which is fed by the vessel and whose needs are independent on the delivering structure.
Then combining Qu\'emada's formula for viscosity with fluid mechanics equations in a vessel, we use the equivalent viscosity $\mu_{eq}$ of our fluid in the vessel. It is defined as the viscosity a Newtonian fluid would need in order to dissipate the same amount of viscous energy than our fluid inside that vessel.
Interestingly when \fah effect can be neglected, the equivalent viscosity is driven only by the mean shear rate in the branch $\gpmoy$ and by the red blood cells volumetric fraction $H_D$: $\mu_{eq}(\gpmoy,H_D) = \frac{G(\gpmoy,H_D)}{8 \gpmoy}$, with $G$ a smooth function.
Explanations about why and how they are such related are detailed in section II-A of supplementary materials. 
\fah effect occurs mainly for vessels whose diameters are smaller than $300 \ \mu m$, see for example \cite{pries_blood_2008} or section III of supplementary materials. For such vessels, the independence of the equivalent viscosity on the vessel radius $r$ is lost. As a consequence, the equivalent viscosity can be expressed using a smooth function $K$: $\mu_{eq}(\gpmoy,r,H_D) = \frac{K(\gpmoy,r,H_D)}{8 \gpmoy}$, see details in section II-B of supplementary materials.

\section{Extending Murray's law to Qu\'emada's fluid model}

In this section, we extend Murray's law to non-Newtonian fluids that can be modeled with Qu\'emada's fluid model. Thus we take into account the dependence of viscosity on shear rates. We start with the assumption of no phase separation effects in the fluid. Next, we study the case where \fah effect occurs in the fluid.

\subsection{Without phase separation effects}

Now we will show how these principles allow to extend Murray's law to any fluid that can be modeled with Qu\'emada's model -and consequently to blood- as long as \fah effects are negligible. Let us consider a fluid that can be modeled with Qu\'emada's model, and let us assume its flow rate $F$ is going through a vessel with radius $r$ and length $l$. The dissipated power $W$ defined by Murray \citep{hess_prinzip_1914,murray_physiological_1926, alarcon_design_2005} divides into two parts: $W = W_H + W_M$, where $W_H$ is the power dissipated by the flow, and $W_M$ the energy consumption rate of the fluid (in the case of blood, this is a metabolic consumption rate). With the above results and denoting $\alpha_b$ the energy consumption rate per unit volume of fluid, we have
\begin{equation}
\label{DP1}
W_H = \frac{8 F^2 \mu_{eq}(\gpmoy) l}{\pi r^4} \ \text{ and } \ W_M = \alpha_b \pi r^2 l 
\end{equation}
The design principle proposed by Murray is to search for a minimum of $W$ relatively to the radius of the vessel, thus solving $\frac{\partial W}{\partial r} = 0$. In our case, expanding this last equality leads to a non linear equation which depends only on $\gpmoy$, see section III of supplementary materials.
As a consequence, the optimal configuration is reached when the mean shear rate in the vessel solves the preceding equation, independently on the fluid flow rate $F$ or the vessel sizes $r$ or $l$. Interestingly, we observed that the optimal shear rates are as small as possible, to get low viscous dissipation, but still high enough to keep viscosity in its lowest values. In the case of blood, the optimal shear rate is $\gpmoy_{noF}^* \sim 12.5 \ s^{-1}$, it can be easily located on figure \ref{viscosityvs}: it stands on the left of the low viscosity plateau (red dashed line).

Let us apply this result to a bifurcation where the flow rate in the parent branch is $F_p$ and the flow rates in the daughter branches are $F_1$ and $F_2$. Their respective radii are denoted $r_p$, $r_1$ and $r_2$. If $\gpmoy^*_{noF}$ is the solution of $\frac{\partial W}{\partial r} = 0$, then $\gpmoy^*_{noF}$ is the optimal mean shear rate in the three vessels and $\gpmoy^*_{noF} = \frac{F_p}{\pi r_p^3} = \frac{F_1}{\pi r_1^3} = \frac{F_2}{\pi r_2^3}$. By flow conservation, we have an additional equation which is $F_P = F_1 + F_2$. Combining these equations finally leads to Murray's original law: $r_p^3 = r_1^3 + r_2^3$. 

The method and these result are most general, they apply to any fluid whose viscosity is monotonously driven by the shear rate in the vessel. 

\subsection{With \fah effect.}

\begin{figure}[h!]
\centering
\includegraphics[height=4.5cm]{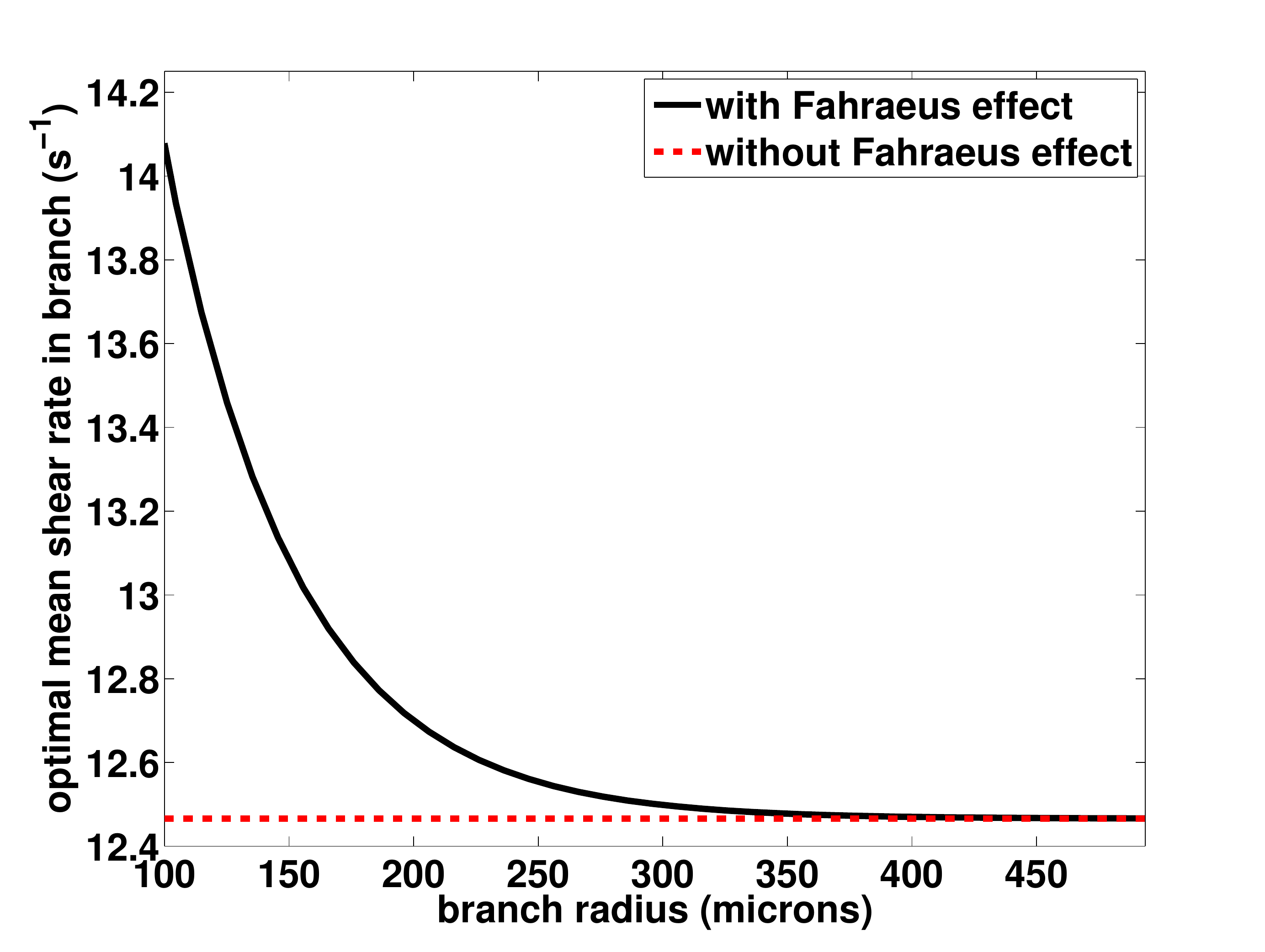}
\caption{In black, the optimal mean shear rate in a branch as a function of the radius of the branch with \fah effect (function $r \rightarrow \gpmoy^*_{noF}(r)$ in text); in red dotted line, the optimal mean shear rate without \fah effect. Blood case: $H_D = 0.45$ and $\alpha_b = 77.8 \ J.m^{-3}.s^{-1}$ \citep{taber_optimization_1998,alarcon_design_2005}.}
\label{gpmoyFPlot}
\end{figure}

The \fah effect in blood is a phase separation effect due to the biphasic characteristics of blood: red blood cells tend to migrate toward the centre of the vessels and blood near the vessel wall is depleted in red blood cells. \fah effect contributes to blood \fahNoS-Lindvquist effect \citep{fahraeus_suspension_1929,fung_biomechanics_1997}. It becomes non negligible for vessels with diameters smaller than $300 \ \mu m$, see for example \citep{pries_blood_2008}. To estimate the role of \fah effects on Murray's law in such vessels, we approximated this effect by assuming that a red-blood-cell depleted layer stands near the wall of the vessels, see section II-B of supplementary materials. The thickness of this depleted layer, and consequently the whole fluid dynamics into that vessel, depend on the branch radius \citep{pries_blood_2008}. As a consequence, the equivalent viscosity in a branch where \fah effect occurs is not anymore dependent on the mean shear rate only, it also becomes dependent on the radius of the branch: $\mu_{eq}(\gpmoy,r) =  \frac{K(\gpmoy,r)}{4 \gpmoy}$, with $K$ a smooth function.  In this case the power dissipation in the fluid used for deducing Murray's law is $W_H = \frac{8 F^2 \mu_{eq}(\gpmoy,r) l}{\pi r^4}$.
The energy consumption rate of the fluid $W_M$ remains $W_M = \alpha_b \pi r^2 l$. As before, the radius that minimizes the total work $W = W_H + W_M$ solves $\frac{\partial W}{\partial r} = 0$.
Expanding this last equation shows that the minimum power is reached on a curve $r \rightarrow \gpmoy^*_F(r)$ that depends on the red blood cells volumetric fraction $H_D$. As an example, we plotted the curve computed in the case of blood on figure (\ref{gpmoyFPlot}). When $r$ is large enough, say larger than about $300 \ \mu m$ \citep{pries_blood_2008}, then the dependence on $r$ is lost and $\gpmoy^*_F(r) = \gpmoy^*_{noF}$.

Let us now consider a bifurcation where the mean shear rate in the parent branch is $\gpmoy_p$ and the mean shear rates in the two daughter branches are $\gpmoy_1$ and $\gpmoy_2$. Their respective radii are denoted $r_p$, $r_1$ and $r_2$. Using flow conservation through the bifurcation, we can deduce that the optimal radii occur for the following equality:
\begin{equation}
\label{murrayLawF}
\gpmoy^*_F(r_p) \ r_p^3 = \gpmoy^*_F(r_1) \ r_1^3 + \gpmoy^*_F(r_2) \ r_2^3
\end{equation}
As expected, when \fah effect vanishes, the previous equation simplifies into the original Murray's law $r_p^3 = r_1^3 + r_2^3$.

\section{Extended Murray's law in fractal trees}

We are now interested in a tree structure built as a cascade of cylinders. The branches divides regularly into $n$ smaller identical branches. The number of divisions between a vessel and the root vessel of the tree defines its generation index. The tree root stands at generation $0$ and the tree limbs end at generation $N$. The size of the branches are defined thanks to the homothety ratio or scaling factor $h$ \citep{mauroy_optimal_2004, mauroy_influence_2010} that corresponds to the relative change in vessels diameter and lengths in a bifurcation, i.e. if $r_i$ is the radius of vessels of generation $i$, then the radii of vessels of generation $i+1$ are $r_{i+1} = h \times r_i$.
It has been shown that bifurcating points in tree structures affects in a complex way the fluid dynamics, even for the Newtonian case, see for example \cite{pedley_energy_1970}. However, for the sake of keeping the model tractable, we will stick to Murray's initial hypothesis and assume that fluid dynamics is not perturbed at the bifurcations and that velocities remain fully developed all along the branches.
Blood flow rate through the tree is assumed constant to mimic the needs of a downstream organ that are independent on the network geometry.
\begin{figure}[h!]
\centering 
\includegraphics[height=5cm]{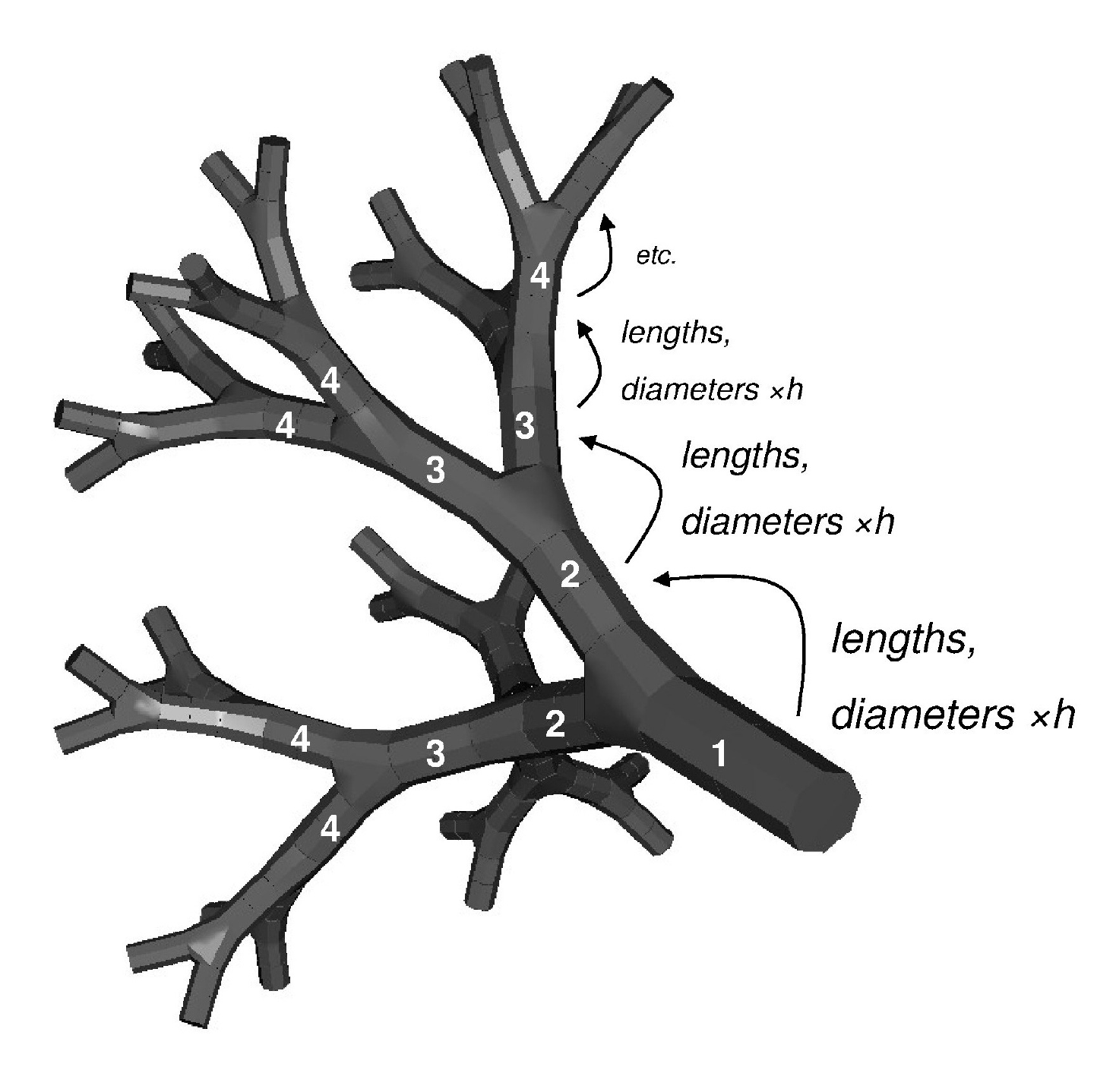}
\caption{Tree network structure with $n=2$ and $N = 6$. The tree is dichotomous and the vessels size decreases at each generation: their diameters and lengths are multiplied by the homothety factor $h<1$ after each bifurcation.}
\label{treegeom}
\end{figure}

\subsection{Mean shear rates}

Denoting $F_{i}$ the blood flow in a branch of generation $i$ and  $S_i = \pi r_i^2$ the surface of its circular cross section, then the mean shear rate in that branch is $\left<\dot{\gamma}_i\right> = \frac{F_{i}/S_i}{r_i}=\frac{F_{i}}{\pi r_i^3}$. Since the tree branches divide into $n$ smaller identical branches, the total blood flow in a branch of generation $i$ is $n$ times the total blood flow in a branch of the next generation $i+1$. Then the mean shear rate in a branch of generation $i+1$ is
$\left<\dot{\gamma}_{i+1}\right> = \frac{F_{i+1}}{\pi r_{i+1}^3} = \frac{F_{i}}{\pi r_i^3} \frac{1}{n h^3} = \left<\dot{\gamma}_{i}\right> \frac{1}{n h^3}$. Thus, the regularity of the structure (fractal) induces also a scaling law on the mean shear rates in the tree. This scaling law depends only on the homothety ratio $h$. Depending on the position of the factor $\frac{1}{n h^3}$ relatively to $1$, the mean shear rate has different behaviors: 
\begin{enumerate}
\item If $h>\left(\frac{1}{n}\right)^{1/3}$, then the mean shear rate decreases along the generations, consequently blood viscosity tends to increase along the generations; 
\item If $h=\left(\frac{1}{n}\right)^{1/3}$, then the mean shear rate remains constant along the generations and so for blood viscosity; 
\item If $h<\left(\frac{1}{n}\right)^{1/3}$, then the mean shear rate increases along the generations, consequently blood viscosity tends to decrease along the generations.
\end{enumerate}
\begin{figure}[h!]
\centering
\includegraphics[height=4.5cm]{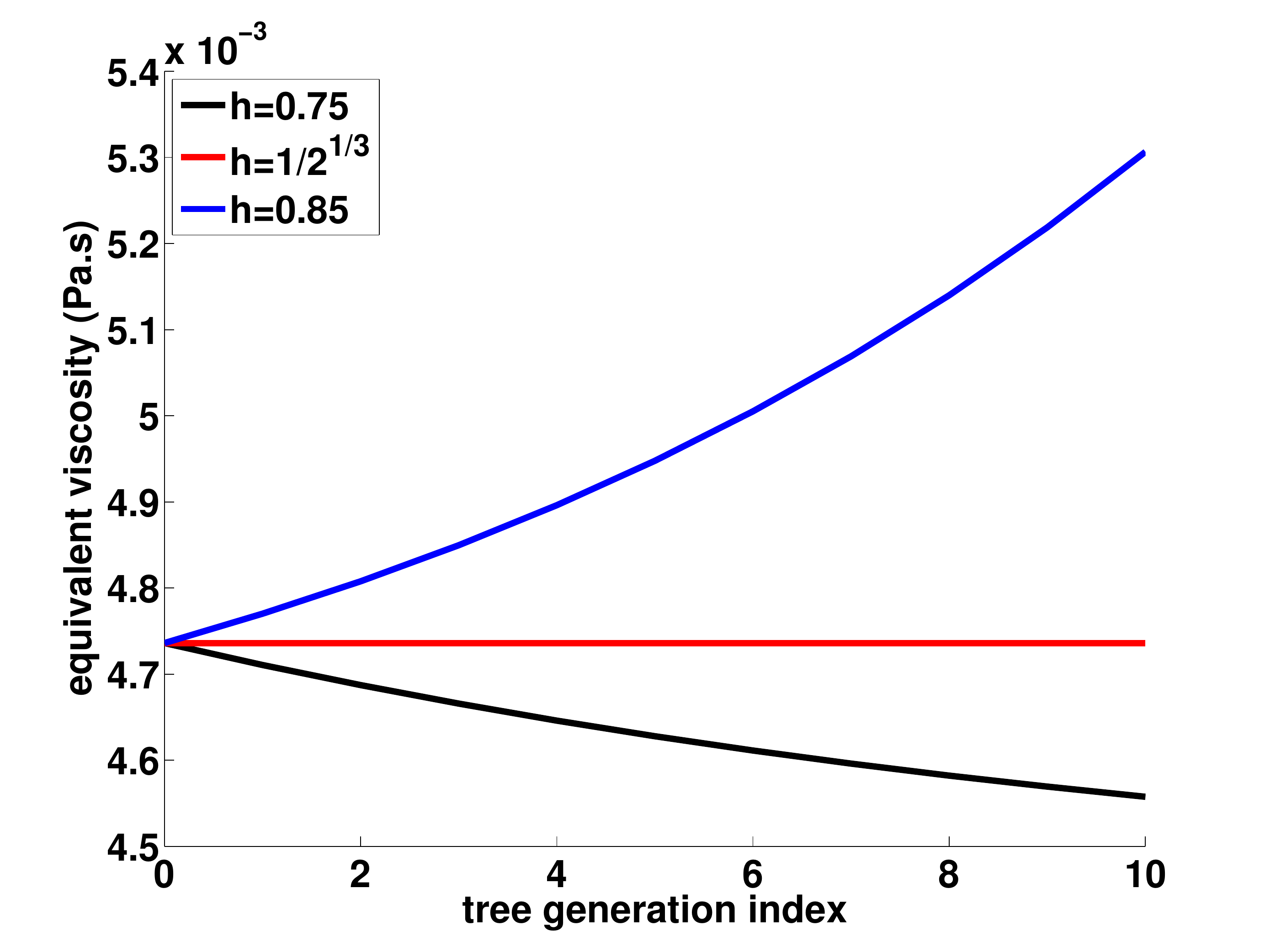}
\caption{Mean viscosity variation in a dichotomous tree ($n=2$) with $N = 10$ generations for three different values of the homothety reduction factor $h$. In the case plotted \fah effect is negligible and the shear rate in the root branch of the tree is $640 \ s^{-1}$.}
\label{viscosityvsGens}
\end{figure}
Blood viscosity depends on shear rate in a non linear way, with a plateau of high viscosities at low shear rates and a plateau of lower viscosities for high shear rates. Between the two plateaus, for medium shear rates, blood viscosity varies steeply, see figure \ref{viscosityvs}. 
When the shear rate decreases or increases along the generations of the tree, then viscosity will vary more strongly if shear rate variation makes the viscosity go through the steep part. Consequently, the amplitude of viscosity variation depends on the initial mean shear rate $\left<\dot{\gamma}_0\right>$ in the root branch of the tree: a high shear rate in the root branch can induce notable viscosity variations throughout the tree only if shear rate is decreasing enough along generations; similarly, a low shear rate in the root branch can induce notable viscosity variations throughout the tree only if the shear rate increases enough along the generations. In any other case, the viscosity is stalled, either in one of the two plateaus or because mean shear rate does not vary much if $h \sim \left(\frac{1}{n}\right)^{1/3}$.

\subsection{Without phase separation effects}

In the absence of \fah effect, Murray's law extended to Qu\'emada's model states that the mean shear rates in the tree branches should be all equal to the optimal mean shear rate $\gpmoy^*$. Since the mean shear rate is following a scaling law in the tree, the only way for the tree to minimize the dissipative power throughout the tree is for the scaling parameter $\frac1{n h^3}$ to be equal to $1$, i.e. $h_* = \left(\frac{1}{n}\right)^\frac13$. This optimal configuration corresponds to mean shear rates and equivalent viscosities being constant throughout the whole tree. This result is fully compatible with blood arterial macrocirculation \citep{pries_blood_2008}. For blood, it has been estimated that $\alpha_b = 77.8 \ J.m^{-3}.s^{-1}$ and $H_D = 0.45$ \citep{taber_optimization_1998, alarcon_design_2005}; with these numbers, the optimal mean shear rate we predict is $\gpmoy^*_{noF} \sim 12.5 \ s^{-1}$, see figure \ref{viscosityvs}. 

Interestingly, except in very large vessels and in microcirculation, blood arterial network geometry exhibits a constant mean shear rate along its generations \citep{pries_blood_2008}, with a homothety ratio of about $\left(\frac12\right)^{\frac13}$ \citep{rossitti_vascular_1993}, as predicted by our model. This raises the question on how physiology can reach this configuration. A scenario proposed by \cite{zamir_optimality_1976} is based on the today well-known fact that endothelial cells standing on the arterial walls are able to respond to shear stress stimuli \citep{pries_microvascular_2005}. If this response is made accordingly to a shear stress threshold common to all cells, increasing the vessel diameter if the shear stress is over a threshold (by cells division) and decreasing the vessel diameter if the shear stress is below the same threshold (either by apoptosis or migration), then the resulting tree will exhibit a homothety ratio that is $\left(\frac12\right)^{\frac13}$. This is a direct consequence of the scaling law on mean shear rates described in the previous section.

\subsection{With \fah effect.}

\begin{figure}[h!]
\centering
\includegraphics[height=4.5cm]{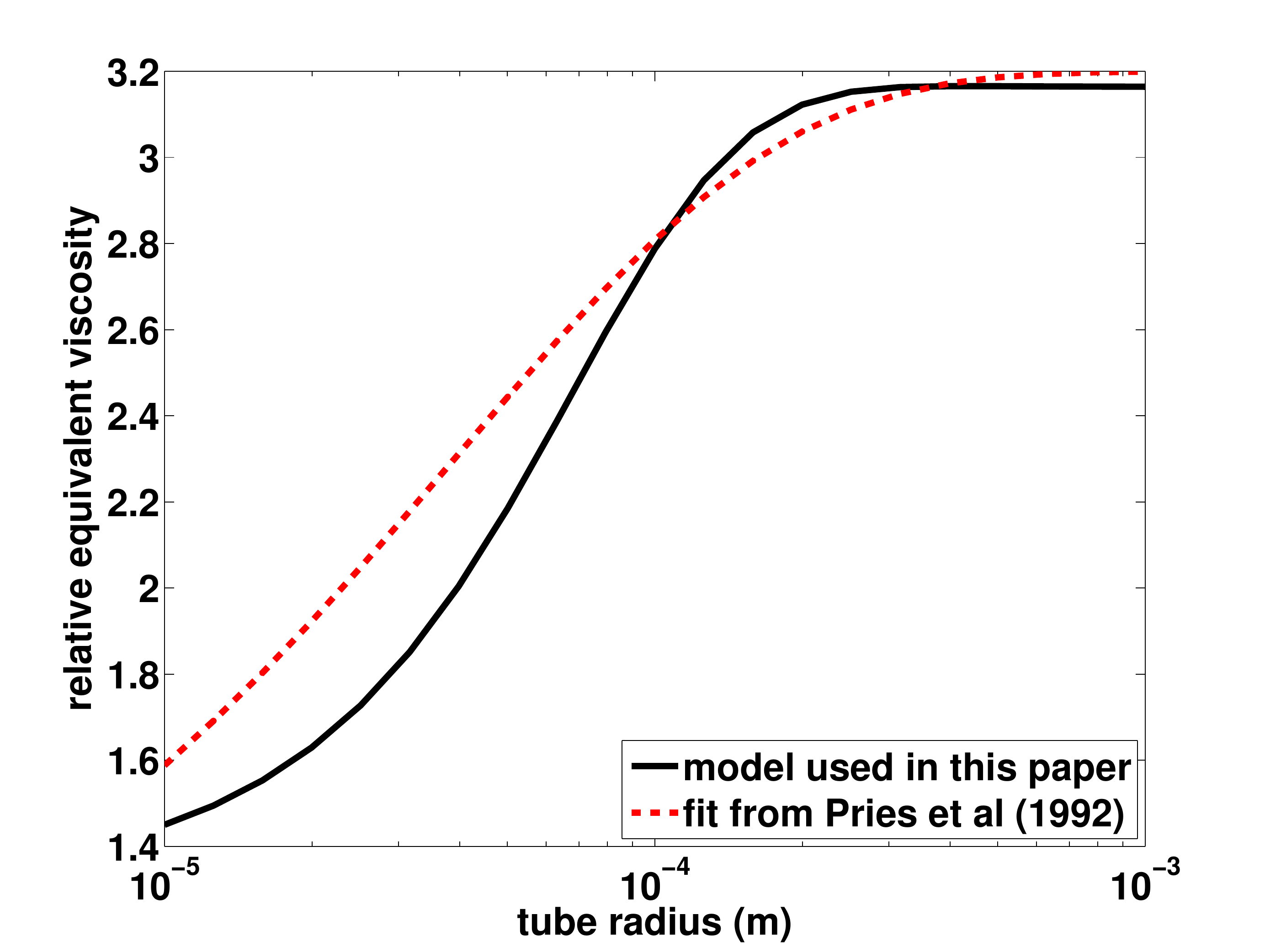}
\caption{Relative viscosity versus tube radius for a constant pressure drop in the tube (high shear rates). The dashed red line represent the mean fit from experiments performed in \cite{pries_blood_1992}. The black continuous line represents the behavior of our model. Notice that we are interested only in having a similar trend and not to exactly fit the red dashed curve.}
\label{relViscosity}
\end{figure}

The case of arterial microcirculation is more complex, because it includes phase separation effects. 
Many experiments have shown the influence of \fah effect on equivalent viscosity, as reported in \cite{pries_blood_1992}. Fully detailed analysis and curve fittings are available in \cite{pries_blood_1992} and \cite{pries_blood_2008} about the relative viscosity, which is the ratio between the equivalent viscosity and the embedding fluid viscosity (plasma in the case of blood). Our model remains in the range of the experimental data and gets a similar trend as that of the data in these studies, as shown on figure \ref{relViscosity}. Notice that by altering the parameters of our model, we would be able to better agree with the fit given in \cite{pries_blood_1992}, but this is out of the scope of this paper.

The optimal configuration of each division in the tree meets the conditions for the extended Murray's law equation (\ref{murrayLawF}) with daughter branches all identical. In addition, the optimal configuration also verifies the scaling law for the branches radii, i.e. $r_{i+1} = h r_i$. These equations make the scaling factor $h$ generation dependent, thus scaling factors need now to be indexed with their generation index $i$. The optimal $h_i$ in term of Murray's design verifies the equation: $n h_{i,*}^3 \gpmoy^*_{F}(h_{i,*} r_{i,*}) = \gpmoy^*_{F}(r_{i,*})$.

\begin{figure}[h!]
\centering 
\includegraphics[height=5cm]{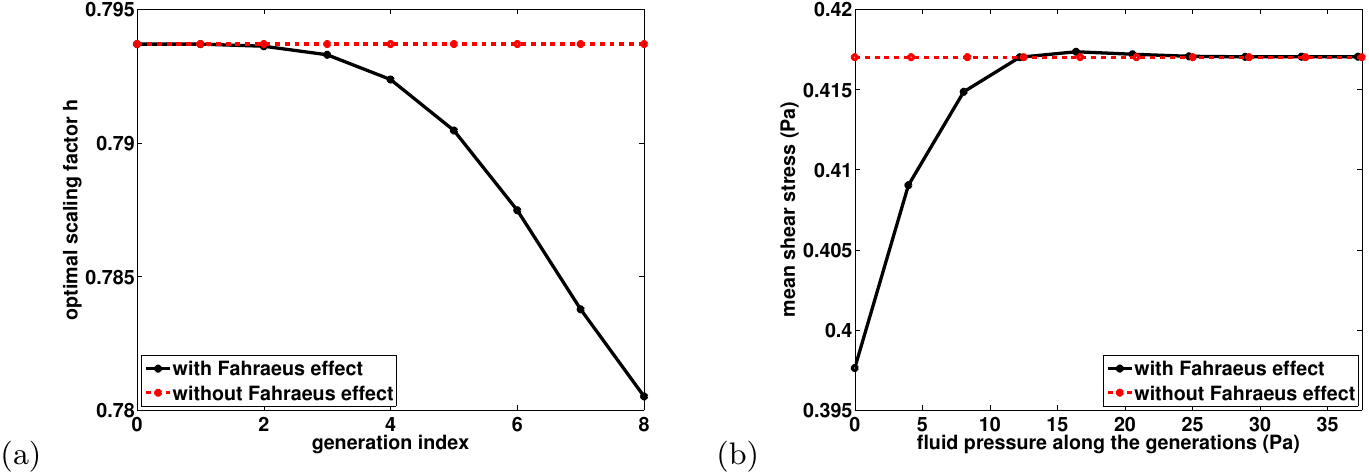}
\caption{Example for blood circulation in a dichotomous tree ($n=2$), with 9 generations, a root radius of $800 \ \mu m$ and a ratio of length over radius for the branches equal to $10$; Blood properties are from \citep{taber_optimization_1998,alarcon_design_2005}: $H_D = 0.45$ and $\alpha_b = 77.8 \ J.m^{-3}.s^{-1}$. {\bf (a)}: Optimal scaling factor with \fah effect (black) and without \fah effect (red).
{\bf (b)}: Mean shear stress versus pressure within the tree. Deepest generations are on the left and upper generations are on the right. 
}
\label{optimhF}
\end{figure}

We computed numerically the optimal scaling factors in the case of blood for a dichotomical tree ($n=2$) with nine generations, see figure \ref{optimhF}A. The root branch radius is $800 \ \mu m$, thus \fah effect is negligible in the first generations. In the optimal configuration, the three first generations are bifurcating with a scaling factor of $\left(\frac{1}{2}\right)^\frac13 \sim 0.7937$, then \fah effect becomes non negligible and it affects the optimal scaling factors of the next generations. 
The function $r \rightarrow \gpmoy^*_{F}(r)$ is a decreasing function, thus $h_{i,*} < \left(\frac{1}{n}\right)^\frac13 = h_*$.  Thus the optimal configuration has tighter branches where \fah effect occurs than the optimal configuration without \fah effect.
Indeed, \fah effect improves the lubrication of the core fluid in the tube because the red blood cell depleted layer near the wall is less viscous that the core layer. Consequently, equivalent viscosities decrease and mean shear rates increase along the generations as soon as \fah effect appears.

The mean shear stress in a branch where \fah effect occurs is 
$\sigma_{i,*} = \frac18 K(\gpmoy^*_{F}(r_{i,*}),r_{i,*})$. Since $K$ is an increasing function of the mean shear rate, the mean shear stress increases along the generations when \fah effect occurs (see section II-B of supplementary materials for details). Because pressure is decreasing along the generation, the mean shear stress is a decreasing function of pressure within the tree, see the example of blood on figure \ref{optimhF}b. This result is in agreement with the observation of arterial blood microcirculation \citep{pries_blood_2008}, 
and confirms the conclusions of \cite{alarcon_design_2005} that \fah effects might be the core phenomena driving the decrease of blood equivalent viscosity and wall shear stresses in the microcirculation.

\section{Conclusion}

In this study, we extended Murray's law to fluids that can be modeled with Qu\'emada's model. In a cylindrical tube, we showed that the mean shear rate drives the behavior of the equivalent viscosity of the tube and thus of the power dissipated in the flow. The consequence is that Murray's optimization principle in a cylindrical tube brings an universal optimal mean shear rate that does not dependent on the flow amplitude or on the size of the tube. In the case of Qu\'emada's law mimicking blood behavior, we found that this optimal mean shear rate is $\gpmoy_{noF}^* \sim 12.5 \ s^{-1}$. This value is in accordance with arterial macrocirculation data where the mean shear rates in the successive vessels remain almost constant at about $10 \ s^{-1}$. Applied to a fractal tree whose branches divide into $n$ identical smaller branches, we found that the optimal tree scaling factor remains the same as for a Newtonian fluid, i.e. $\left(\frac1{n}\right)^{\frac13}$.

However, if the vessel diameter is small enough, \fah phase separation effect occurs. A red blood cells depleted layer appears near the wall of the vessel, thus making the equivalent viscosity dependent on the mean shear rate and on the vessel radius $r$. We derived a new optimal configuration from Murray's optimisation principle, which is expressed through a decreasing function that relates the optimal mean shear rate in a vessel with its radius: $r \rightarrow \gpmoy^*_F(r)$. For large $r$, this function maps the optimal mean shear rate without \fah effect. The function behavior relates to the lubrication phenomenon induced by \fah effect. We also derived the optimal configuration of a fractal tree when \fah effect occurs. We showed that the optimal scaling factors become dependent on the generation index, on the size of the root of the tree and on the function $r \rightarrow \gpmoy^*_F(r)$, i.e. $n h_{i,*}^3 \gpmoy^*_{F}(h_{i,*} r_{i,*}) = \gpmoy^*_{F}(r_{i,*})$. The optimal configuration of the fractal tree allows tighter branches when \fah effect occurs and induces a drop in mean shear stresses along the tree, which has been reported in the literature for blood arterial microcirculation.

Our study is based on the sole fact that viscosity is a monotonous function of the fluid shear rate. Consequently, our results hold not only for fluid that can be modeled with Qu\'emada's model, but also for any fluid for which this monotonous condition is true. This make our study most general and apply for example to shear-thinning or shear-thickening fluids. 
Concerning blood, interestingly, the optimal configuration fits that of large circulation. Large circulation is thus in a state where non-Newtonian effects remain small, just on the edge of the steep part of blood rheogram, as shown on figure \ref{viscosityvs}. It is important to notice that this property is a result of the optimization process and not an hypothesis. This result can only be obtained by including the non Newtonian behavior of blood in the optimization process. This also highlights why, although it was based on Newtonian hypothesis, the original Murray's law fitted nevertheless blood large circulation.
Of course, many other phenomena might play a role in the complex behavior of arterial blood circulation. 
Hence, our model does not account for the complex regulation processes that exist in blood network. These regulation processes are likely to affect the optimal configuration and enhance the robustness of the system.
In large circulation, the role of inertia, turbulence or oscillating flow caused by the heart beat may influence Murray's optimal design. In the microcirculation, the roles of other phase separation effects have to be taken into account to improve the validity of the predictions, for example the role of the glycocalyx molecules standing on the vessels walls \citep{pries_endothelial_2000}. Also, because the power dissipation in Murray's optimal design is not symmetric relatively to its optimal value, biological noise affecting blood vessel sizes may also influence the optimal configuration \citep{mauroy_influence_2010,vercken_dont_2012}. The inclusion of such a noise in Murray's optimal design may bring interesting insights on blood vessels size distribution in the arterial network.
Finally, blood network is regulated and this regulation probably affects the optimal configuration.

\section*{Acknowledgments}
The authors would like to thank Philippe Dantan, Patrice Flaud and Daniel Qu\'emada (University Paris Diderot - Paris 7, Paris, France) for very fruitful discussions and support about this work.

\newpage

\noindent{\Large{\bf{Appendix}}}

\appendix

\section{Qu\'emada's law}
\label{quemadaVar}

The local  viscosity $\mu$ is a function of both local shear rate $\gp(s)$ and local hematocrit $H(s)$ as stated by Qu\'emada's law:

$$
\mu(\dot{\gamma},H) = \mu_p \left( 1 - \frac{H}{H_{\infty}(H)} \right)^{-2} \left( \frac{1+k(H)}{\chi(H)+k(H)} \right)^2
$$

\noindent with $k(H)=\left( t_c(H) |\dot{\gamma}| \right)^\frac12$, $\chi(H) = \left( 1-\frac{H}{H_0(H)} \right) / \left( 1-\frac{H}{H_\infty(H)} \right)$; $\mu_p$ is the viscosity of plasma and $\mu_p=1.6 \ 10^{-3} \ Pa.s$. The quantities $t_c$ (seconds), $H_0$ and $H_\infty$ are functions of the red blood cells volumetric fraction $H$. We use the fits from \citep{cokelet_rheology_1987}:

$$
t_c(H) = e^{6.1508-27.293 H+25.60 H^2-3.697 H^3} 
$$

$$
H_{\infty}(H) = 2 \ e^{-1.3435+2.803 H-2.711 H^2+0.6479 H^3} 
$$

$$
H_0(H) = 2 \ e^{-3.8740+10.410 H-13.80 H^2+6.738 H^3} 
$$

The formulas for $t_c$, $H_{\infty}$ and $H_0$ come from data fitting, they are valid for red blood cells volumetric fraction up to $0.7$ \citep{comolet_biomecanique_1984} and for moderate shear rates, i.e. larger than about $0.1 \ s^{-1}$. Other fitting methods using second order polynomials can be used, see for example \citep{sriram_non-newtonian_2014}. Using such alternative fit induces very few differences on the results predicted in this work. 

\section{Computation of the effective viscosity}
\label{computemueq}

We assume the vessels to be cylindric and the fluid to be axi-symmetric and fully developed in all the vessels. Following Qu\'emada's law, the viscosity $\mu(\dot{\gamma}(s),H(s))$ of the fluid at a radial position $s$ of a vessel depends on the local shear rate $\dot{\gamma}(s)$ and on the local red blood cell concentration $H(s)$. The pressure drop per unit length is denoted $C$, it is assumed independent on the position. In this frame, the fluid dynamics reduce to an equation on the shear rate $\gp(s)$ at radial position $s \in [0, r]$: $\mu\left(\gp,H\right) \gp = \frac{C s}{2}$.

\begin{figure}[h!]
\centering
A
\includegraphics[height=4.5cm]{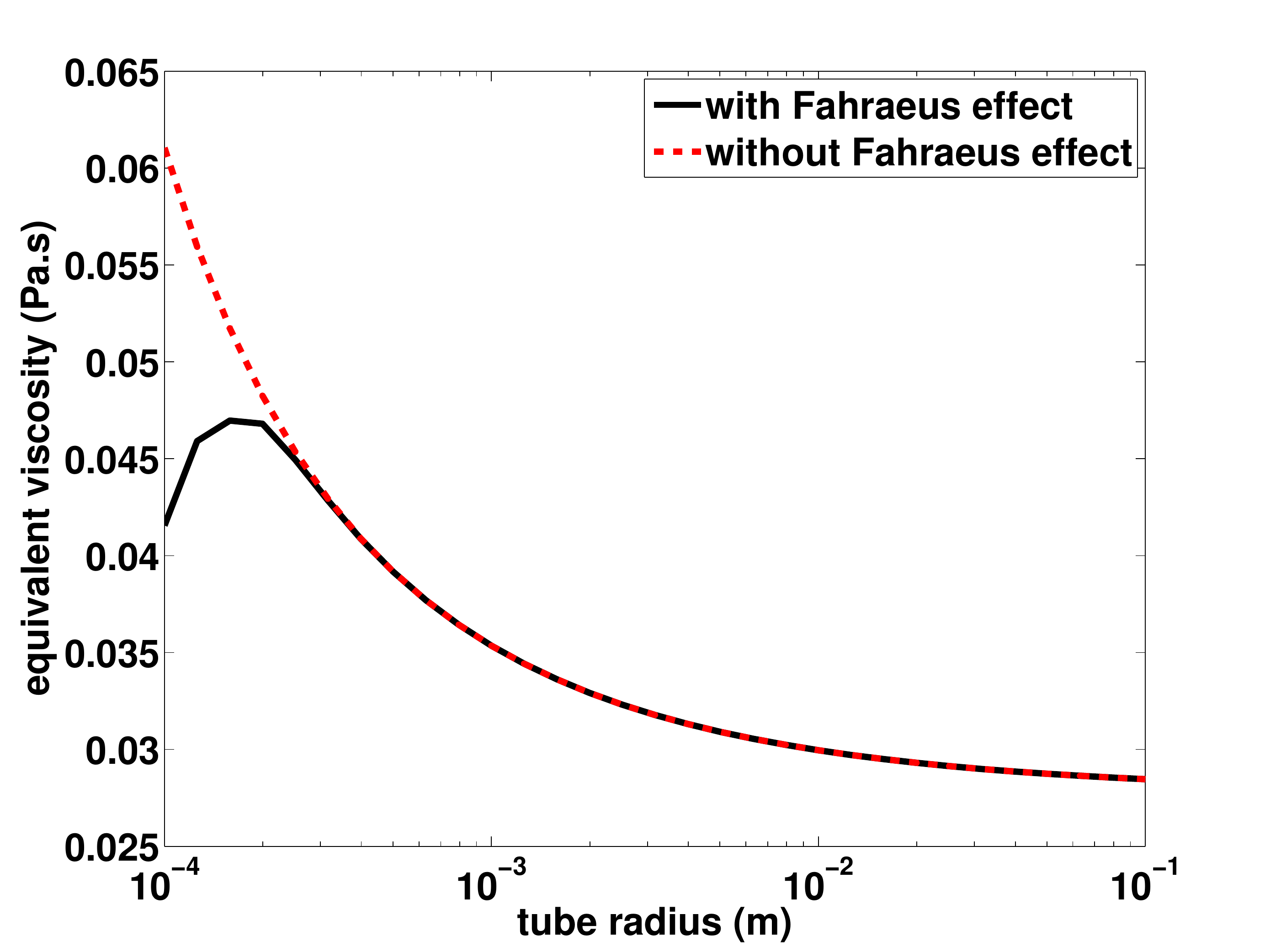}
B
\includegraphics[height=4.5cm]{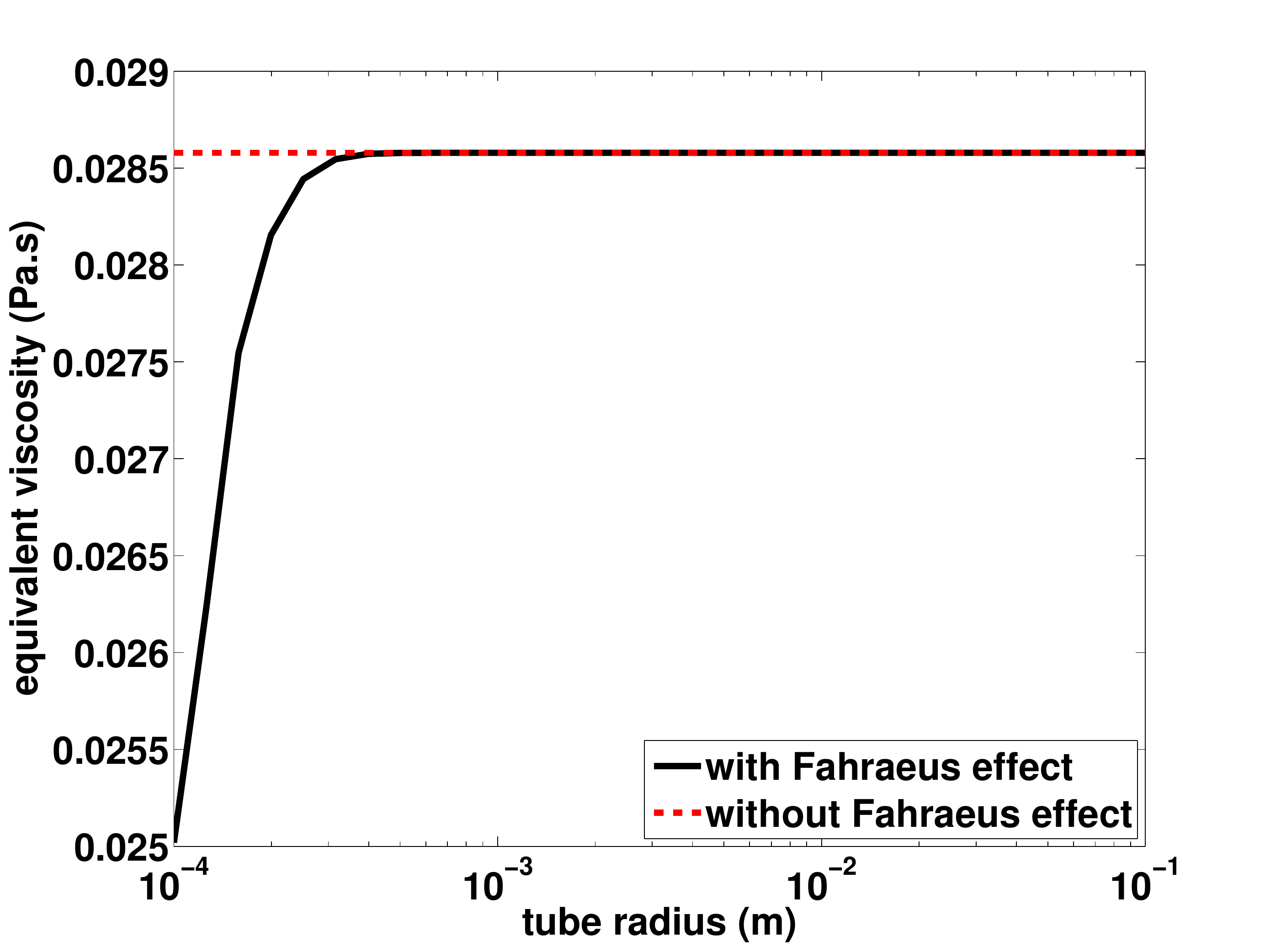}
\newline
C
\includegraphics[height=4.5cm]{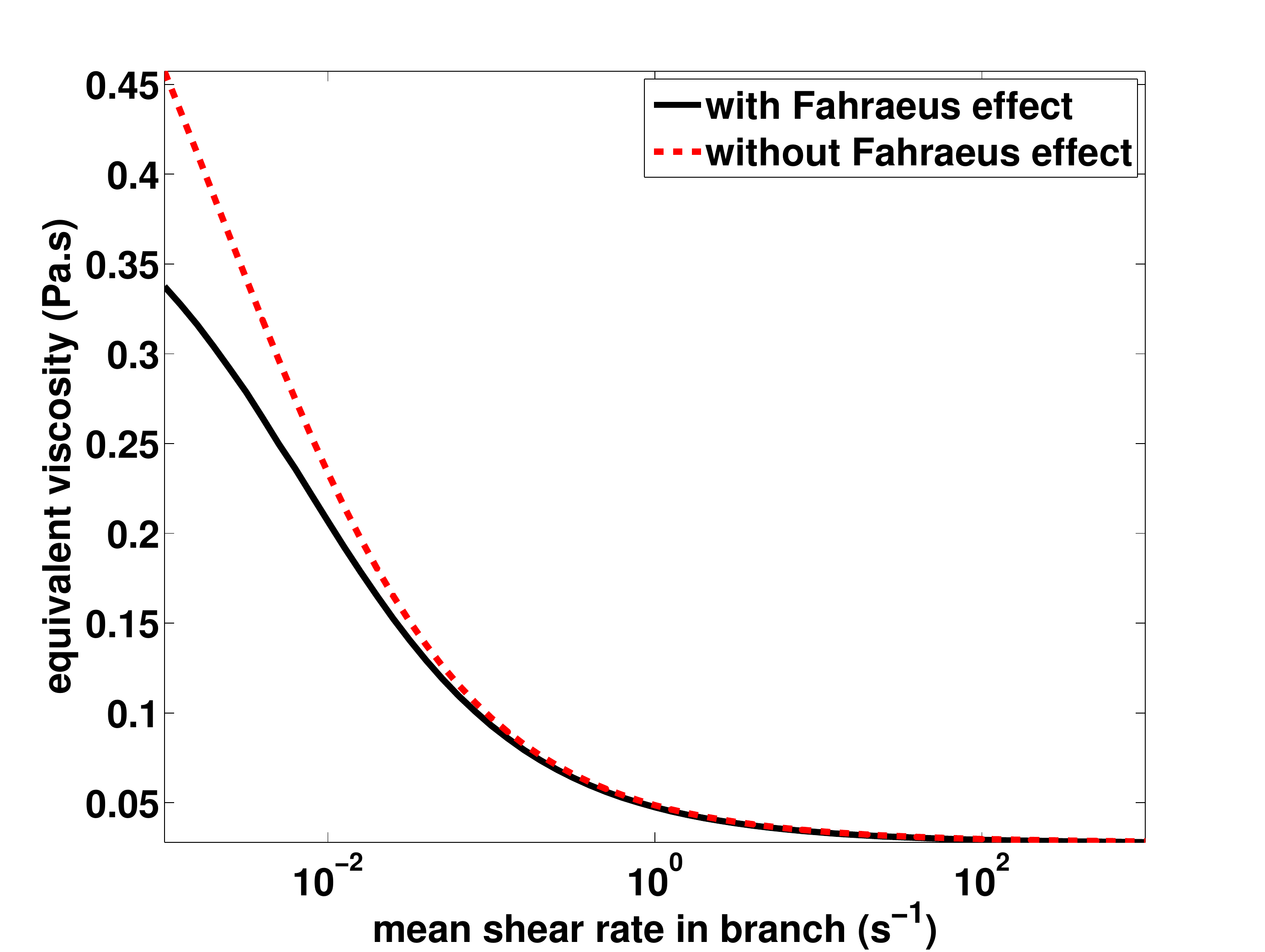}
\caption{A: \fah effect plays a role on branch with small radii only, typically smaller than $300 \ \mu m$. B: Effective viscosity $\mu_{eq}$ depends on mean shear rate in branch unless \fah effect occurs, case with mean shear rate of $2/\pi \times 10^{-3} \ s^{-1}$. C: Dependence of effective viscosity $\mu_{eq}$ with mean shear rate (example with tube radius $200 \ \mu m$ where \fah effect occurs and plays a role mainly at low mean shear rates ($< 0.1 \ s^{-1}$)).}
\label{FvsNoF}
\end{figure}

Once the expression of the viscosity $\mu$ is known, this equation shows that the shear rate at radial position $s$, $\gp$, is a function of $C s$ and $H$ only. Using Qu\'emada's law, it is possible to reach an analytical expression for the shear rate $\gp$, and consequently for the flow rate $F$ in the branch as a function of the radius of the branch $r$, of the pressure drop per unit length $C$ and of the discharge hematocrit $H_D$, see appendix \ref{computemueq} for the details. Note that for branches whose radius is sufficiently large, say larger than $300 \mu m$, \fah effect is negligible and the blood hematocrit $H$ in the branch is homogeneous and equal to the discharge hematocrit $H_D$, see figure \ref{FvsNoF}. From these relations, and the expression of the hydrodynamical resistance of a cylindrical tube of length $L$, $R = CL/F = (8 \mu L)/(\pi r^4)$, we can compute the effective viscosity $\mu_{eq}$ of the fluid in a branch depending on the quantity $C r$ in the branch and the red blood cells volumetric fraction $H_D$ in the branch: $\mu_{eq}(r, C ,H_D) = \frac{\pi r^4 C}{8 F(r,C,H_D)} =\frac{C r}{8 \gpmoy(r,C,H_D)}$.

\subsection{Without \fah effect.}
\label{withoutF}

In this section, we consider a cylindrical branch with radius $r$ and length $L$. Blood flow in the branch is assumed to be fully developed and axi-symmetrical. The quantity $C$ is the pressure drop per unit length in a vessel, it is assumed to be a constant in the whole vessel. We do not take \fah effect into account, ths red blood cells volume fraction $H$ is assumed constant in the whole vessel. The shear rate at radial position $s$ is then denoted $\gp(s,C,H)$ and the fluid dynamics in the branch with these hypotheses reduces to
and is the solution of the following equation:

\begin{equation}
\label{FDeq}
\mu\left(\dot{\gamma}(s,C,H),H(s)\right) \dot{\gamma}(s,C,H) = \frac{C s}{2}
\end{equation}

The first consequence of this equality is that the shear rate $\gp(s,C,H)$ can actually be seen as a function of $Cs$ and $H$ only : $\gp(Cs,H)$. The quantity $C s$ is a constraint; denoting $x$ the axis of the branch, $2 \pi C s dx$ represents the infinitesimal  increase of the pressure force felt by a particle of fluid whose radial position in the branch is $s$ and which moves a distance $dx$ forward into the branch.

 The second consequence of equation (\ref{FDeq}) is that we can compute an analytical solution for the local shear rate $\gp(C s,H)$:

\begin{equation}
\label{gpsa}
\gp(C s,H) = \left(\frac{\sqrt{\gp_1(C s,H)}}{2}-\frac{\sqrt{\gp_3(H)}}{2}+\frac{\sqrt{\left(\sqrt{\gp_1(C s,H)}-\sqrt{\gp_3(H)}\right)^2+4 \sqrt{\gp_1(C s,H) \gp_2(H)} }}{2}\right)^2
\end{equation}

\noindent with $\gp_1(C s,H) = \frac{|C s|}{2 \mu_{P}} \left(1-\frac{H}{H_{\infty}(H)}\right)^2$, $\gp_2(H) = \frac{\chi(H)^2}{{t_c(H)}}$ and $\gp_3(H) =  \frac{1}{{t_c(H)}}$. Then the flow $F$ in a branch is:

\begin{equation}
F = - 2 \pi \int_0^r \gp(C s,H) \frac{s^2}{2} ds
\end{equation}

Consequently, the mean shear rate in the branch $\gpmoy$ is a function of $Cr$ and $H$, as shown by the following equalities:

\begin{equation}
\label{gpmoyVSCr}
\gpmoy = \frac{F}{\pi r^3} = - \frac{\int_0^r \gp(C s,H) s^2 ds}{r^3} =  - \frac{\int_0^{Cr} \gp(y,H) \frac{y^2}{C^2} \frac1C dy}{r^3} =  - \frac{\int_0^{Cr} \gp(y,H) y^2 dy}{\left(Cr\right)^3} = \gpmoy(C r, H)
\end{equation}

Moreover, the mean shear rate $\gpmoy$ is a strictly increasing function of the pressure drop per unit length (and as a matter of fact of the radius of the branch too, since it depends on $Cr$), thus there exists a function $G$ such that $C r = G(\gpmoy,H)$ that relates the mean shear rate $\gpmoy$ to $C r$ and $H$. No analytical expression for $G$ is available but it can be easily computed numerically by inverting equation (\ref{gpmoyVSCr}).

From these results, we can get an analytical formula of the effective viscosity of the fluid in a branch as a function of $C r$ and $H$:

\begin{equation}
\label{mueqCr}
\mu_{eq}(C r, H) = \frac{\pi r^4 C}{8 F} = \frac1{\gpmoy(C r, H)} \frac{C r}{8} = - \frac{C r^4}{16} \left( \int_0^r \gp\left(C s,H\right) \frac{s^2}{2} ds \right)^{-1}
\end{equation}

or using $\gpmoy$ as the variable:

\begin{equation}
\label{mueqgpmoy}
\mu_{eq}(\gpmoy, H) = \frac18 \frac{G(\gpmoy,H)}{\gpmoy}
\end{equation}

To summarize, when \fah effect can be neglected, the quantity $C r$ is a function of the mean shear rate in the branch $\gpmoy$ and of the red blood cells volumetric fraction $H_D$, they are related through the function $G$: $C r = G(\gpmoy,H_D)$. 
Thus, when \fah effect does not occur, then the effective viscosity is driven by the mean shear rate in the branch $\gpmoy$ and red blood cells volumetric fraction $H_D$: $\mu_{eq}(\gpmoy,H_D) = \frac{G(\gpmoy,H_D)}{8 \gpmoy}$. 

\subsection{With \fah effect.}
\label{withF}

In this section, we consider a cylindrical branch with radius $r$ and length $l$. Blood flow in the branch is assumed to be fully developed and axi-symmetrical. The quantity $C$ is the pressure drop per unit length in a vessel, it is assumed to be a constant in the whole vessel. Blood hematocrit is not anymore assumed constant in the whole vessel since we account for \fah effect in this section.\\

\noindent{\it Red blood cells distribution in a vessel.}

\begin{figure}[h!]
\centering 
\includegraphics[height=2cm]{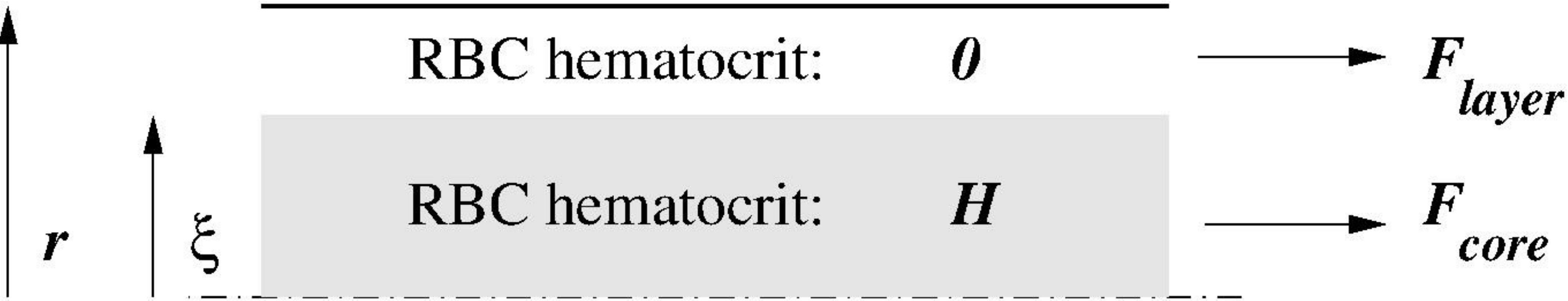}
\caption{Scheme of the blood flow inside an axi-symmetric vessel.}
\label{vesselscheme}
\end{figure}

We assume that red blood cells stand only in the core of the vessel where the hematocrit is constant and equal to $H$ (core hematocrit). The radius of the core is $\xi \leq r$. Plasma flows in the layer near the wall (its flow is $F_{layer}$). We neglect the role of the glycocalyx layer \citep{pries_endothelial_2000} since its thickness, of about $1$ micron, is small relatively to the diameters of the vessels considered, that are larger than $50$ microns. Plasma and red blood cells flow in the core (their flow is $F_{core}$), see figure \ref{vesselscheme}.

The red blood cells depletion near the vessel wall induces the \fah effect that corresponds to a decrease of the mean hematocrit in the vessel relatively to the discharge hematocrit $H_D$ in large vessels. Since the blood that circulates into the vessel comes from larger vessels, by flow conservation \citep{fung_biomechanics_1997, pries_blood_2008} we can relate discharge hematocrit $H_D$ to core hematocrit $H$

\begin{equation}
\label{HDvsH}
H_D = \frac{H(r,C,H_D) \ F_{core}(r,C,H_D)}{F_{core}(r,C,H_D)+F_{layer}(r,C,H_D)}
\end{equation}

The dependence of the mean red blood cells volumetric fraction in a vessel, called tube hematocrit $H_T$, with tube radius and discharge hematocrit is well documented \citep{pries_blood_2008}: if the radius of the tube $r$ is measured in microns, then

\begin{equation}
\label{HTvsHD}
H_T(r, H_D) = H_D \times \left( H_D+(1-H_D)\times (1+1.7 e^{-0.415 \times 2r} - 0.6 e^{-0.011 \times 2r}\right)
\end{equation}

Finally, tube hematocrit is related to core hematocrit by $H_T(r, H_D) \pi r^2 = H(r,C,H_D) \pi \xi(r,C,H_D)^2$ and

\begin{equation}
H_T(r,H_D)=H(r,C,H_D) \ \frac{\xi(r,C,H_D)^2}{r^2}
\label{tubeH}
\end{equation}

We can thus define the following function, which represents how red blood cells volumetric fraction varies along the radius of the vessel:

\begin{equation}
\label{Hvss}
\left\{
\begin{array}{ll}
\text{if } 0 \leq s \leq \xi \text{ (core), }& H(s) = H(r,C,H_D) = \frac{r^2}{\xi(r,C,H_D)^2} H_T(H_D,2r)\\
\text{if } \xi < s \leq r \text{ (wall layer), }& H(s) = 0\\
\end{array}
\right.
\end{equation}

In order to be able to compute the values of $H(r,C,H_D)$ and $\xi(r,C,H_D)$, we need to add to this set of equations the equations of fluid dynamics.\\

\noindent{\it Fluid mechanics.}

The shear rate $\gp$ at radial position $s$ is the solution of the following equation:

\begin{equation}
\label{FDeqF}
\mu\left(\dot{\gamma}(s,C,H(s)),H(s)\right) \dot{\gamma}(s,C,H(s)) = \frac{C s}{2}
\end{equation}

As in the case with no \fah effect, this shows that shear rate at radial position $s$ is a function of $C s$ and $H(s)$. However, on the contrary of the case without \fah effect, the dependence of $\gp$ on $s$ remains through the dependence of $H(s)$ on $s$. From equation (\ref{FDeqF}), we can compute an analytical solution for the local shear rate $\gp(C s,H(s))$:

\begin{equation}
\label{gpsaFah}
\left\{
\begin{array}{l}
\text{if } 0 \leq s \leq \xi \text{ (core), }\\
\ \ \ \ \gp(C s,H(s)) = \left(\frac{\sqrt{\gp_1(C s,H(s))}}{2}-\frac{\sqrt{\gp_3(H(s))}}{2}+\frac{\sqrt{\left(\sqrt{\gp_1(C s,H(s))}-\sqrt{\gp_3(H(s))}\right)^2+4 \sqrt{\gp_1(C s,H(s)) \gp_2(H(s))} }}{2}\right)^2\\
\text{if } \xi < s \leq r \text{ (wall layer), } \ \gp(C s,H(s)) = \frac{|C s|}{2 \mu_p}\\
\end{array}
\right.
\end{equation}

\noindent with $\gp_1(C s,H(s)) = \frac{|C s|}{2 \mu_{P}} \left(1-\frac{H(s)}{H_{\infty}(H(s))}\right)^2$, $\gp_2(H(s)) = \frac{\chi(H(s))^2}{{t_c(H(s))}}$ and $\gp_3(H(s)) =  \frac{1}{{t_c(H(s))}}$. 

As a consequence of the dependence of the shear rate on $s$, the mean shear rate and the effective viscosity in the branch are both dependent on the radius of the branch $r$:

\begin{equation}
\label{gpmoyF}
\gpmoy(r,C,H_D) = \frac{F_{core}(r,C,H_D) + F_{layer}(r,C,H_D)}{\pi r^4}
\end{equation}

\begin{equation}
\mu_{eq}(r,C,H_D) = \frac{C r}{8 \gpmoy(r,C,H_D)}
\end{equation}

Once $r$ and $H_D$ are fixed, the mean shear rate $\gpmoy$ in the branch is a strictly increasing function of the pressure drop per unit length $C$, thus there exists a function $K$ such that $C r = K(\gpmoy,r,H_D)$. The function $K$ is computed numerically (see next section). Thus, effective viscosity can be reformulated into new variables:
 
\begin{equation}
\mu_{eq}(\gpmoy,r,H_D) = \frac{K(\gpmoy,r,H_D)}{8 \gpmoy}
\end{equation}

\noindent{\it Solving the system.}

In this section, we first explain how the diameter of the red blood cells core $\xi$ and the volumetric fraction of red blood cells in the core, $H$, are computed once we know the radius of the branch $r$, the pressure drop per unit length in the branch $C$ and the discharge hematocrit $H_D$. Then we explain how we obtain the function $k$ that relates $C$ to the mean shear rate in the branch $\gpmoy$.

The red blood cells core $\xi$ and the volumetric fraction of red blood cells in the core $H$ are the solution of the following system that regroup the conservation equations in the branch and the equations from fluid dynamics (see previous sections):

\begin{equation}
\label{sys}
\left\{
\begin{array}{ll}
H = H_D \frac{F_{core}(r,C,H_D) + F_{layer}(r,C,H_D)}{F_{core}(r,C,H_D)} & \text{(equation (\ref{HDvsH}))}\\
\xi^2 = \frac{H_T(r,H_D)}{H} r^2& \text{(equations (\ref{HTvsHD}) and (\ref{tubeH}))}\\
F_{core}(r,C,H_D) = 2 \pi \left( v(\xi) \frac{\xi^2}{2} -\int_0^\xi \dot{\gamma}(Cs,H(s)) \frac{s^2}{2} ds \right) &\\
F_{layer}(r,C,H_D) = - 2 \pi \left( \int_\xi^r \dot{\gamma}(Cs,H(s)) \frac{s^2}{2} ds + v(\xi) \frac{\xi^2}{2}\right)&\\
v(\xi)=-\frac{C}{4 \mu_p} \left( r^2-\xi^2 \right)&\\
\text{$\gp(Cs,H(s))$ is given by equation (\ref{gpsa})}&
\end{array}
\right.
\end{equation}

\noindent The system (\ref{sys}) is normalized and solved using Matlab, thanks to the {\it fsolve} command (based on a Newton method). To compute the two integrals in equation (\ref{sys}), we use the {\it quad} command (adaptative Simpson quadrature).

The mean shear rate in the branch $\gpmoy(Cr,H_D)$ is then computed using equation (\ref{gpmoyF}). The function $k$ that relates the quantity $Cr$ to the mean shear rate $\gpmoy$ is then computed numerically by numerical inversion of the function $\gpmoy(Cr,H_D)$.\\

To summarize, when \fah effect occurs, the results computed without \fah effects remain a good approximation if the vessel radius is larger than about $300 \ \mu m$. For smaller vessels, we lose the independence on the radius of the branch $r$, and the effective viscosity can be expressed using a function $K$ relating $C r$ to $\gpmoy$:
$\mu_{eq}(\gpmoy,r,H_D) = \frac{C r}{8 \gpmoy} = \frac{K(\gpmoy,r,H_D)}{8 \gpmoy}$. In general, $H_D$ is constant and it will dropped off the equations for clarity reasons. All our results would nevertheless depends on the value of $H_D$.

Finally, the mean shear stress in a branch is $\sigma_{i,*} = \mu_{eq}(\gpmoy^*_{F}(r_{i,*}),r_{i,*}) \gpmoy^*_{F}(r_{i,*}) = \frac18 K(\gpmoy^*_{F}(r_{i,*}),r_{i,*})$.

\section{Optimisation of Murray's cost.}
\label{murrayOpt}

The numerical example we gave in this work for optimal mean shear rate use Qu\'emada's fluid and the set of parameters corresponding to blood (see appendix \ref{quemadaVar}). They were computed with $H_D = 0.45$ and $\alpha_b = 77.8 \ J.m^{-3}.s^{-1}$ \citep{taber_optimization_1998,alarcon_design_2005}. The numerical method depends on whether \fah effects occur or not. The first step is always to build the function relating the effective viscosity to mean shear rate, hematocrit and, in the case where \fah effect occurs, with vessel radius, see appendices \ref{computemueq}.

When \fah effect does not occur, then we solved directly the equation $\frac{1}{\pi r l} \frac{\partial W}{\partial r} = 0$, i.e.
$$
\frac{1}{\pi r l} \frac{\partial W}{\partial r} = - 8 \gpmoy^2 \left( 4 \mu\left( \gpmoy \right) + 3 \gpmoy \frac{\partial \mu_{eq}}{\partial \gpmoy}\left( \gpmoy \right) \right) + 2 \alpha_b= 0
$$
\noindent This formula shows that the minimum of $W$ is reached for a given value of $\gpmoy^*_{noF}$. The derivative $\frac{\partial \mu_{eq}}{\partial \gpmoy}$ is computed numerically using a first order method.

When \fah effect occurs, $W$ is given by
$$
\frac{W}{\pi l} = \frac{8 F^2 \mu_{eq}(\gpmoy,r)}{\pi^2 r^4} + \alpha_b r^2
$$
and its derivative of $W$ relatively to $r$ expresses as
$$
\frac{1}{\pi r l} \frac{\partial W}{\partial r} = - 8 \gpmoy^2 \left( 4 \mu\left( \gpmoy \right) + 3 \gpmoy \frac{\partial \mu_{eq}}{\partial \gpmoy}\left( \gpmoy \right) \right) + 2 \alpha_b= 0
$$
This formula shows that the minimum power is reached on a curve $r \rightarrow \gpmoy^*_F(r)$. In the case with \fah effect, we did not work with $\frac{\partial W}{\partial r}$ but instead we minimized directly the cost $W = W_H + W_M$ using a gradient method. A value for the fluid flow rate $F$ is chosen as stated in Murray's optimal design, then we compute the optimal radius $r^*(F)$. Finally, by spanning the whole range of interesting fluid flow rates $F$, we can compute the function relating the mean shear rate in the vessel to the vessel radius with the formula:
$\gpmoy^*(r^*(F)) = \frac{F}{\pi \left(r^*(F)\right)^3}$.

\newpage

\bibliographystyle{apalike}
\bibliography{biblio}

\end{document}